\documentclass[runningheads]{llncs}
\usepackage[T1]{fontenc}
\usepackage{graphicx,verbatim}
\usepackage{amsmath} 
\usepackage{amssymb} 
\usepackage{tabularx}
\usepackage{multirow}
\usepackage{array}
\usepackage{subcaption}
\usepackage{graphicx}
\usepackage{colortbl}
\begin{document}
\title{LVPNet: A Latent-variable-based Prediction-driven End-to-end Framework for Lossless Compression of Medical Images}
\titlerunning{LVPNet}
%

\author{Chenyue Song\textsuperscript{1}, Chen Hui\textsuperscript{1,2,}\thanks{Chen Hui and Shengping Zhang are both corresponding authors.\\This work was supported in part by the Central Guidance for Local Science and Technology Development Fund Projects under grant 2024ZYD0266, in part by the Startup Foundation for Introducing Talent of Nanjing University of Information Science and Technology under grant  2025r029.}, Qing Lin\textsuperscript{3}, Wei Zhang\textsuperscript{1}, Siqiao Li\textsuperscript{1}, Haiqi Zhu\textsuperscript{1}, Shengping Zhang\textsuperscript{1,\(\star\)}, Zhixuan Li\textsuperscript{4}, Shaohui Liu\textsuperscript{1}, Feng Jiang\textsuperscript{2}, Xiang Li\textsuperscript{1}}  
\authorrunning{Chenyue Song et al.}
\institute{\textsuperscript{1}Harbin Institute of Technology, China\\
\textsuperscript{2}Nanjing University of Information Science and Technology, China\\
\textsuperscript{3}Dalian University of Technology, China\\
\textsuperscript{4}Nanyang Technological University, Singapore\\
    \email{cysong@stu.hit.edu.cn, chui@nuist.edu.cn, s.zhang@hit.edu.cn}\\}

\maketitle              
\begin{abstract}
Autoregressive Initial Bits is a framework that integrates sub-image autoregression and latent variable modeling, demonstrating its advantages in lossless medical image compression. However, in existing methods, the image segmentation process leads to an even distribution of latent variable information across each sub-image, which in turn causes posterior collapse and inefficient utilization of latent variables. To deal with these issues, we propose a prediction-based end-to-end lossless medical image compression method named LVPNet, leveraging global latent variables to predict pixel values and encoding predicted probabilities for lossless compression. Specifically, we introduce the Global Multi-scale Sensing Module (GMSM), which extracts compact and informative latent representations from the entire image, effectively capturing spatial dependencies within the latent space. Furthermore, to mitigate the information loss introduced during quantization, we propose the Quantization Compensation Module (QCM), which learns the distribution of quantization errors and refines the quantized features to compensate for quantization loss. Extensive experiments on challenging benchmarks demonstrate that our method achieves superior compression efficiency compared to state-of-the-art lossless image compression approaches, while maintaining competitive inference speed. The code is at \url{https://github.com/scy-Jackel/LVPNet}.


\end{abstract}
\section{Introduction}
In medical imaging applications, the handling, transmission, and backup of data often necessitate the storage of lossless images, resulting in substantial storage demands\cite{b38,b37,b39，b50,b51}.   Lossless image compression techniques play a crucial role in mitigating storage costs by leveraging inherent image correlations to reduce file sizes while maintaining the integrity of the original content. Shannon's source coding theorem \cite{b1} defines the lower bound for code length based on image entropy, which is determined by the image's inherent distribution. Thus, reducing redundant information is a fundamental strategy for image compression. \cite{b17} introduce the use of the LZ77 algorithm \cite{b18} and Huffman coding \cite{b19} to eliminate redundant information. Later, reference \cite{b20} propose the discrete wavelet transform as an effective method for capturing multi-scale image features.

With the advancement of deep learning, various models \cite{b2,b40,b41} have been widely adopted for lossless image compression. Representative methods include autoregressive models (ARMs) \cite{b3,b4,b33}, variational autoencoders (VAEs) \cite{b5,b6,b7,b8,b34,b35}, normalizing flows (NFs) \cite{b9,b10,b11,b12}, prediction model\cite{b21,b22,b23,b36}, as well as diffusion models \cite{b13,b14}. These methods require balancing compression performance with inference speed. For example, ARMs and diffusion models offer impressive compression performance, but come with the trade-off of significantly longer inference times \cite{b15,b16,b44}. Moreover, image compression encompasses both dataset-level and single-image compression. However, existing lossless compression methods are typically tailored for specific data types, facing significant challenges in addressing both application scenarios simultaneously. For instance, while NFs excel in dataset compression, they exhibit relatively weaker performance in single-image compression. Additionally, as the size of the autoregressive prior increases in ARMs, the risk of posterior collapse along the sub-image sequence becomes more pronounced \cite{b25,b42,b43}, which negatively affects compression performance.

To deal with these challenges, we propose LVPNet, a latent-variable-based prediction-driven end-to-end lossless compression framework specially designed for medical images, suitable for both datasets and single images.     By extracting effective low-dimensional latent representations from the entire image, our approach efficiently captures spatial dependencies within the latent space and predicts pixel values in this domain.     Specially, we introduce two key components: First, Global Multi-scale Sensing Module (GMSM) captures a variety of multi-scale features during downsampling and encodes relevant information into the latent variables; Then Quantization Compensation Module (QCM) learns the distribution of quantization errors to compensate for the loss and enhance pixel prediction accuracy.
Our contributions are summarized as follows:
\begin{itemize}
	\item[$\bullet$] We propose latent-variable-based prediction-driven end-to-end framework for lossless compression of medical images, applicable to both dataset and single-image compression tasks.
	\item[$\bullet$] We propose the GMSM and QCM modules to extract more comprehensive information from multi-level features, which helps mitigate the issue of posterior collapse and reduces the information loss from quantization.
	\item[$\bullet$] The proposed model exhibits exceptional inference speed and achieves strong compression performance across benchmark datasets for medical imaging.
\end{itemize}
\begin{figure}[t]
\includegraphics[width=\textwidth]{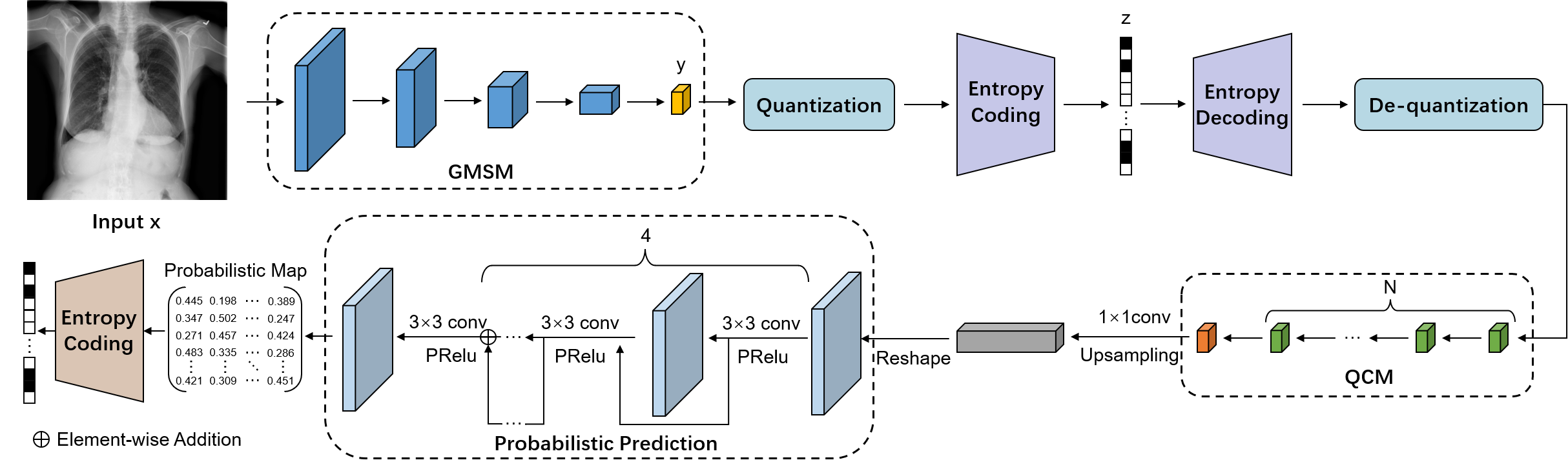}
\caption{Overview of the proposed framework. The framework consists of five distinct modules: a global multi-level Sensing module, quantization, entropy coding, a quantization compensation module, and a probability prediction module.} \label{struct}
\end{figure}
\begin{figure}[t]
    \centering
    \begin{subfigure}[b]{0.44\textwidth}
        \centering
        \includegraphics[width=\textwidth]{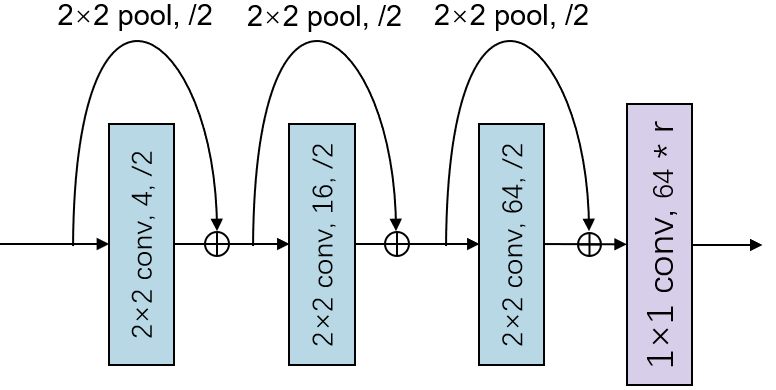}
        \caption{Architecture of GMSM.}
        \label{fig:subfig1}
    \end{subfigure}
    \hfill
    \begin{subfigure}[b]{0.52\textwidth}
        \centering
        \includegraphics[width=\textwidth]{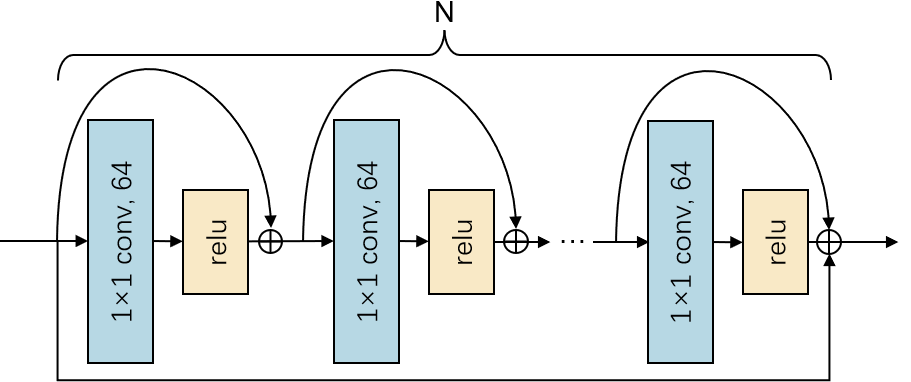}
        \caption{Architecture of QCM.}
        \label{fig:subfig2}
    \end{subfigure}
    \caption{The structure of GMSM and QCM. (a) $r$ is the pre-defined sampling rate. (b) We employ N residual connections and a global connection to propagate features from various layers to the final output.}
    \label{fig:fig}
\end{figure}
\section{Methodology}
\subsection{Overview of LVPNet}
The proposed framework is shown in Fig. \ref{struct}. The input image $x$ is first processed by the global multi-level perception module $G$ to generate the compressed coefficients $y = G(x)$, with a sampling ratio of $r$. $y$ is quantized to map it to the integer domain, producing the latent variable $z = Q(y)$, which is then entropy encoded into a bitstream for storage. Afterward, $z$ undergoes dequantization to produce $\hat{y} = Deq(z)$, and the quantization loss is compensated by the QCM. The compensated features are upsampled and reconstructed, after which the pixel values for the entire image are predicted using the probability prediction module. The process from $z$ to predicted probabilities is denoted as $p_{\theta}(x|z)$, where $\theta$ represents the model parameters. Finally, the predicted probabilities are entropy encoded into a bitstream for storage. LVPNet only needs to store the entropy-encoded $z$, the bitstream of predicted probabilities $p_{\theta}(x|z)$, and the model parameters $\theta$. Since $\theta$ has a fixed size and the bitstream of $z$ remains relatively stable due to the fixed sampling ratio $r$, LVPNet focuses on reducing the size of the entropy-encoded bitstream for $p_{\theta}(x|z)$. 
%
During inference, the latent variable $z$ is decoded, and the predicted pixel values are generated using $p_{\theta}(x|z)$.   By comparing these predictions with the entropy-decoded probabilities corresponding to the ground truth pixel values, the image can be losslessly reconstructed.
\subsection{Global Multi-scale Sensing Module}
Convolutional neural networks hierarchically extract features, where layers proximal to the input capture low-level features such as edges and basic textures, while deeper layers encode abstract and semantic features. Guided by these principles, we propose the GMSM method, as illustrated in Figure \ref{struct}. The framework operates in two stages. In the first stage, $2\times2$ convolutional layers are employed to extract features. In the second stage, features across all hierarchical levels are aggregated and sampled using $1\times1$ convolutional layers, instead of exclusively relying on low-level or high-level features for sampling.

In GMSM, to aggregate multi-level features for sampling, we draw inspiration from ResNet \cite{b26}, employing skip connections to propagate features from various layers to the final stage. Additionally, pooling layers are incorporated to ensure dimensionality alignment. The corresponding formula is expressed as follows:
\begin{equation}\label{f}
f_{t+1}=Conv(f_t)+Pool(f_t)
\end{equation}
where $f_t$ and $f_{t+1}$ denote the sampling features of the $t$-th and $(t+1)$-th layers, respectively.   When transforming the current feature $f_t\in \mathbb{R}^{C\times H\times W}$ into the next-layer feature $f_{t+1}\in \mathbb{R}^{4C\times \frac{H}{2}\times \frac{W}{2}}$, the feature volume remains consistent.   Downsampling is performed through the $1\times1$ convolution layer in the second stage, which facilitates adaptation to various sampling ratios.
\subsection{Quantization and Entropy Coding}
\subsubsection{Quantization}
The downsampled values $y$ are quantized to produce the discrete latent variables $z$, and the inverse operation is applied during dequantization to recover the reconstructed coefficient $\hat{y}$:
\begin{equation}
z=Q(y)=\lfloor y/Q_{step}\rfloor, \hat{y}=Deq(z)=z\times Q_{step}
\end{equation}
where $Q_{step}$ denotes the quantization step size. Due to quantization causing the gradients to become zero, preventing effective backpropagation to the GMSM, we store the gradient prior to the floor operation during forward propagation and use this stored gradient for backpropagation updates. The gradient update formula for the loss function $\mathcal{L}$ in the quantization module is as follows:
\begin{equation}
\frac{\partial \mathcal{L}}{\partial y} = \frac{\partial \mathcal{L}}{\partial z} \cdot \frac{\partial z}{\partial y} \approx \frac{\partial \mathcal{L}}{\partial z} \cdot \frac{1}{Q_{\text{step}}}
\end{equation}
\subsubsection{Entropy Coding}
We apply entropy encoding to both the latent variable $z$ and the predicted probabilities of the pixel values in $p_{\theta}(\hat{x}|z)$. A well-trained model typically results in highly sparse residual latent variables. To encode these sparse residuals, we use Huffman encoding \cite{b27}, while the predicted probabilities are encoded using arithmetic coding \cite{b28}.
\subsection{Quantization Compensation Module}\label{QCM}
The quantization compensation module helps mitigate quantization loss and enhance the predicted probabilities. We employ a basic CNN architecture for the compensation module, as illustrated in Figure \ref{struct}. This network incorporates a global shortcut and multiple residual blocks, which accelerate training. During quantization, a floor operation is applied, causing the dequantized measurement $\hat{y}$ to be slightly smaller than the original measurement $y$. To address this, we use the $relu(\cdot)$ activation function in the final layer of the compensation module, ensuring that the compensation value remains non-negative.
\begin{figure*}[t]\centering
    \begin{minipage}[t]{0.13\textwidth}
        \centering
        \includegraphics[width=\textwidth]{10/00000010_000.png}
    \end{minipage}
    \begin{minipage}[t]{0.13\textwidth}
        \centering
        \includegraphics[width=\textwidth]{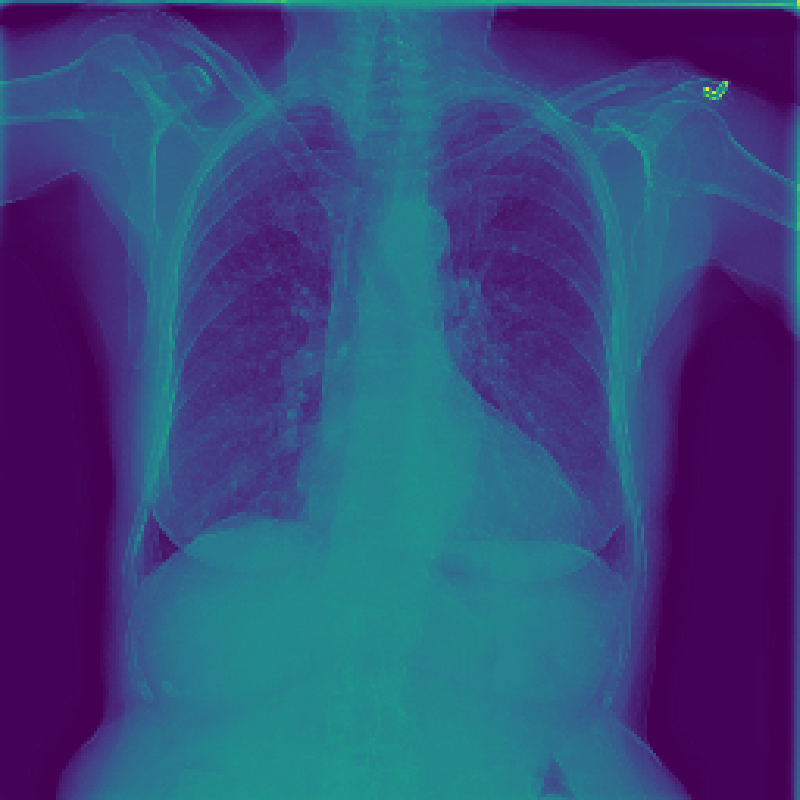}
    \end{minipage}
    \begin{minipage}[t]{0.13\textwidth}
        \centering
        \includegraphics[width=\textwidth]{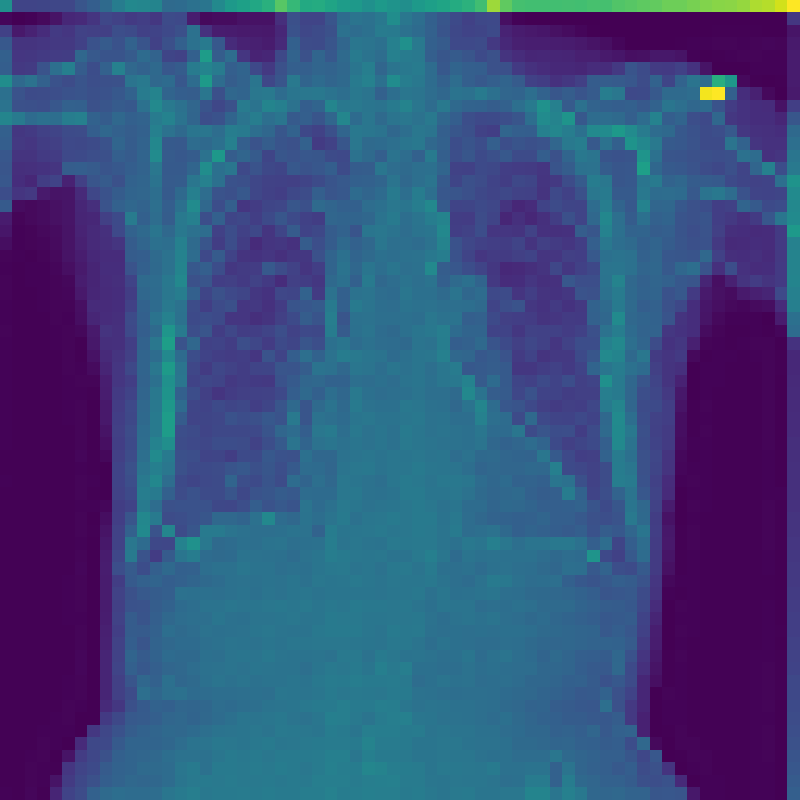}
    \end{minipage}
    \begin{minipage}[t]{0.13\textwidth}
        \centering
        \includegraphics[width=\textwidth]{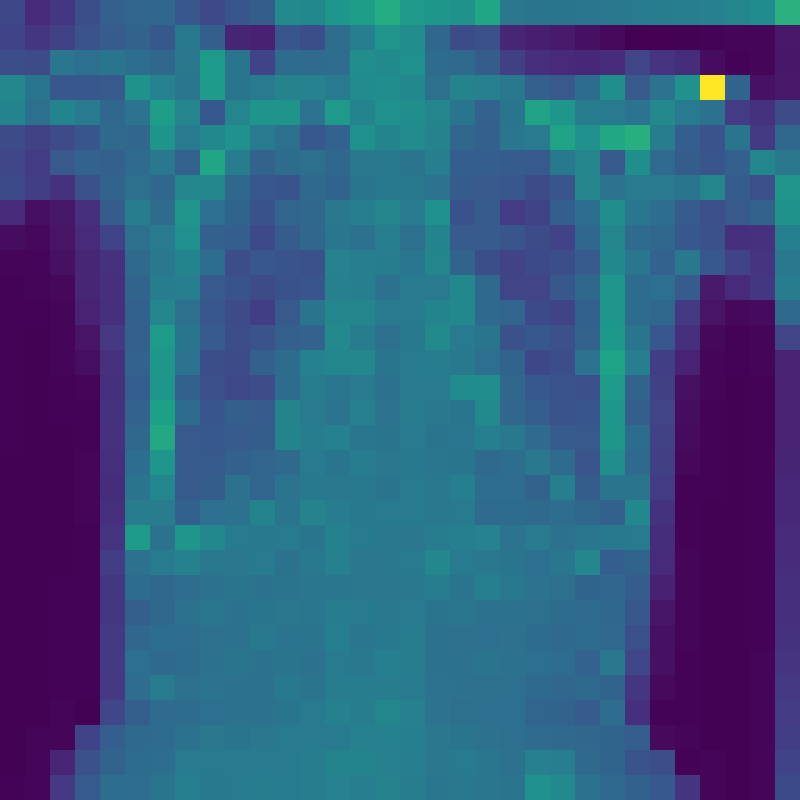}
    \end{minipage}
    \begin{minipage}[t]{0.13\textwidth}
        \centering
        \includegraphics[width=\textwidth]{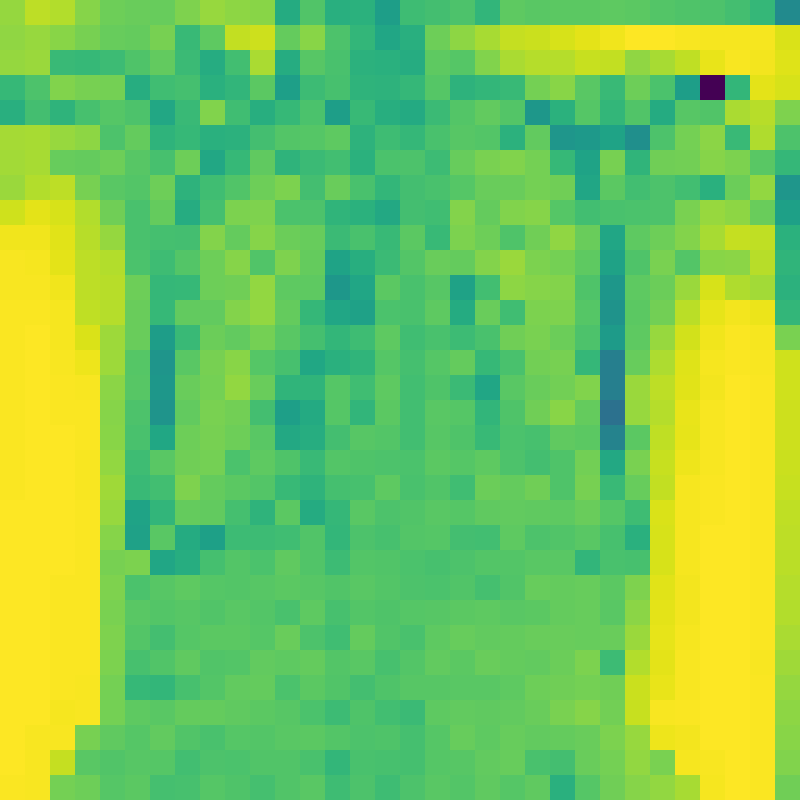}
    \end{minipage}
    \begin{minipage}[t]{0.13\textwidth}
        \centering
        \includegraphics[width=\textwidth]{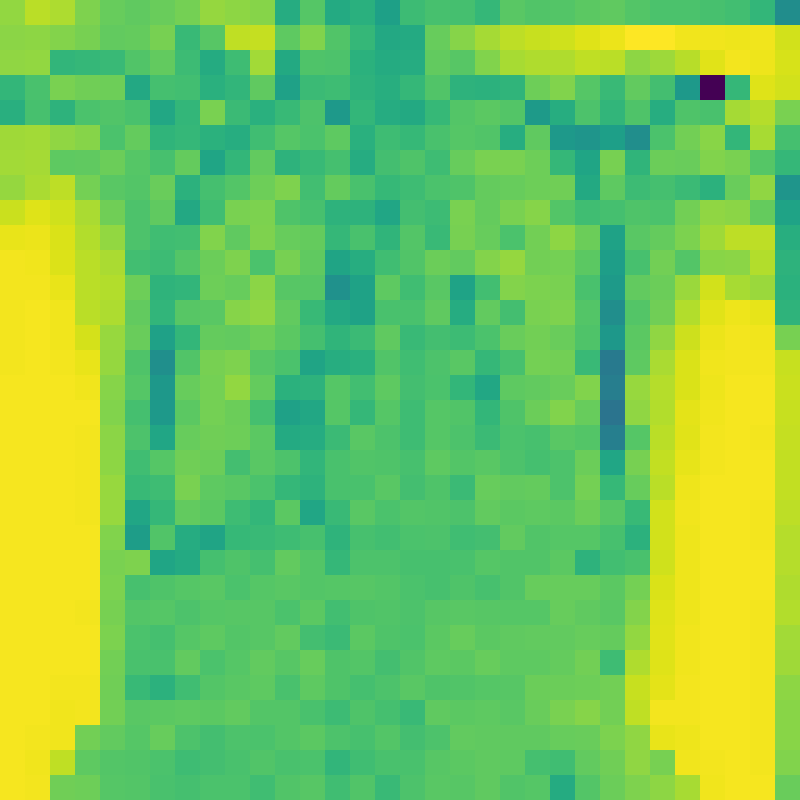}
    \end{minipage}
    \begin{minipage}[t]{0.13\textwidth}
        \centering
        \includegraphics[width=\textwidth]{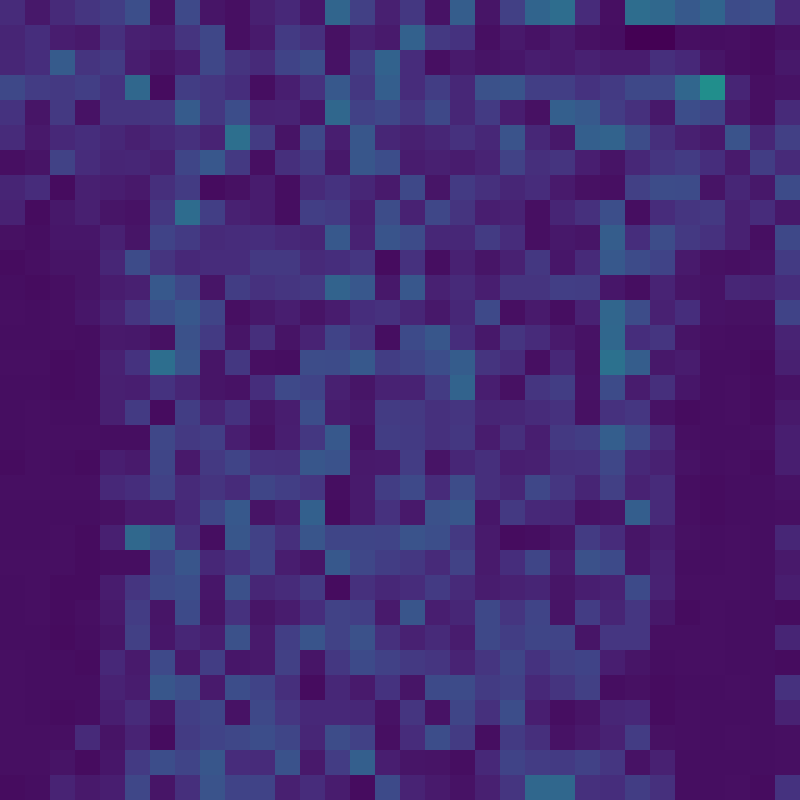}
    \end{minipage}
    \vskip 0.07cm
    \begin{minipage}[t]{0.13\textwidth}
        \centering
        \includegraphics[width=\textwidth]{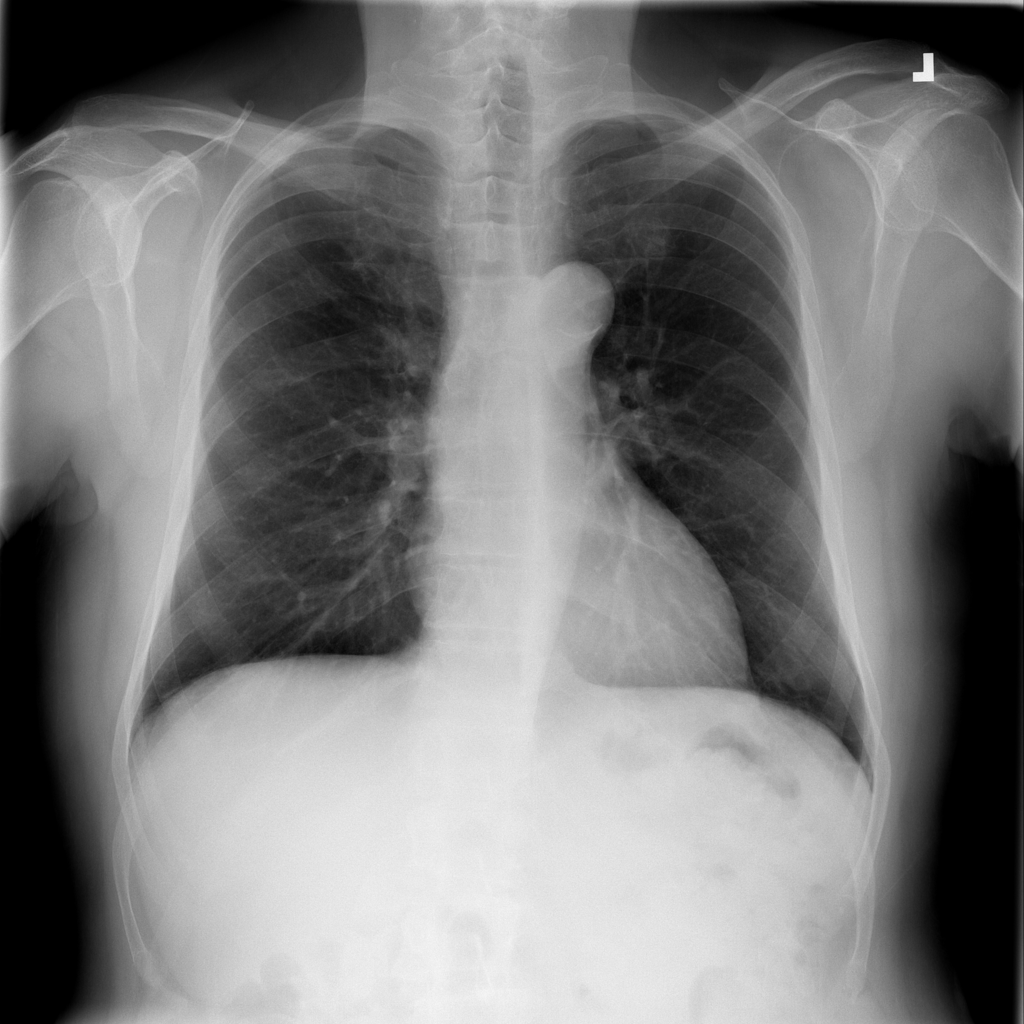}
    \end{minipage}
    \begin{minipage}[t]{0.13\textwidth}
        \centering
        \includegraphics[width=\textwidth]{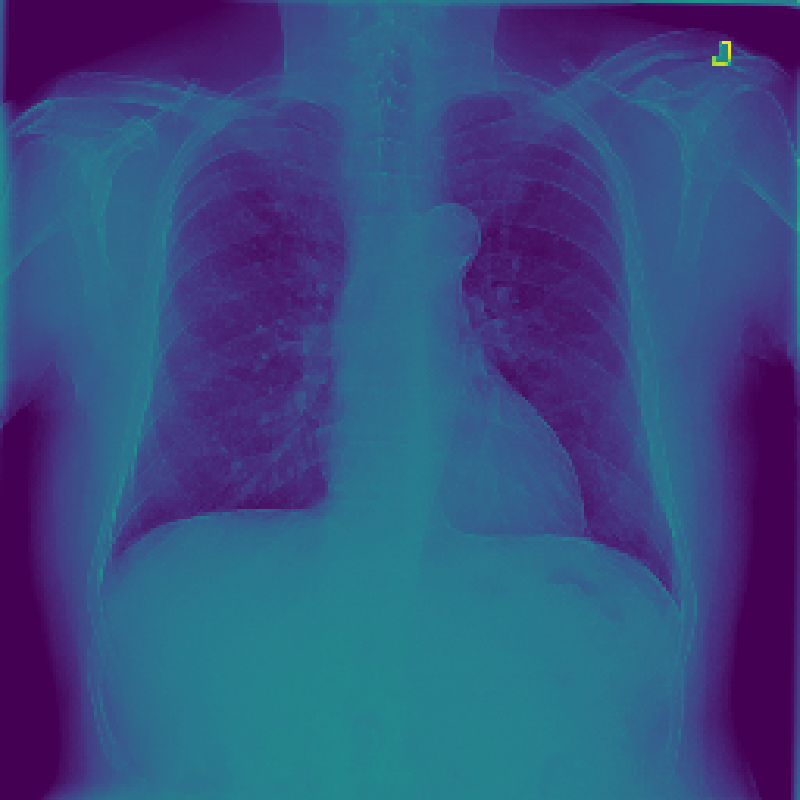}
    \end{minipage}
    \begin{minipage}[t]{0.13\textwidth}
        \centering
        \includegraphics[width=\textwidth]{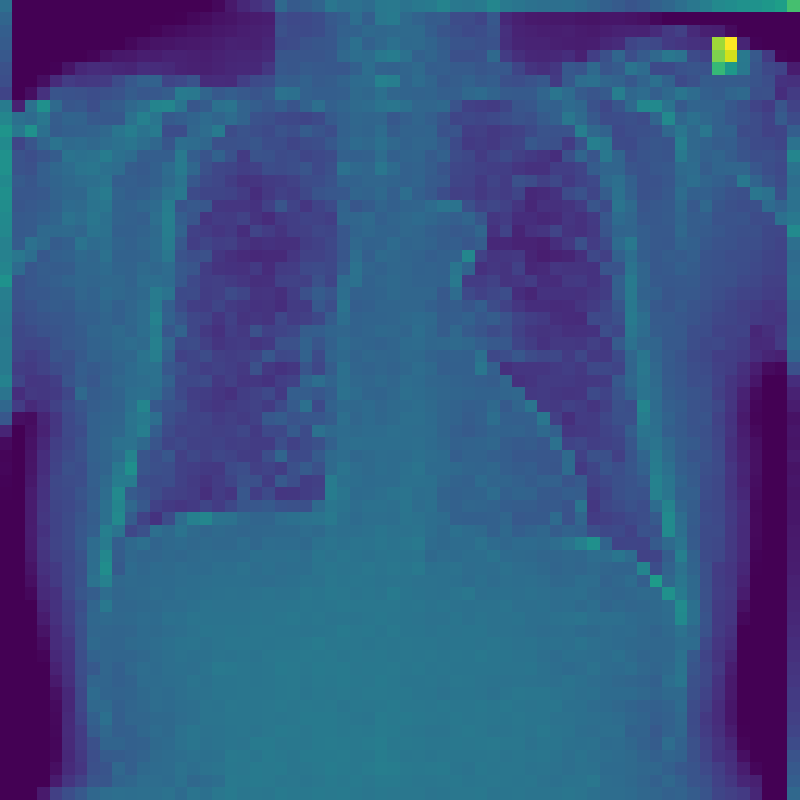}
    \end{minipage}
    \begin{minipage}[t]{0.13\textwidth}
        \centering
        \includegraphics[width=\textwidth]{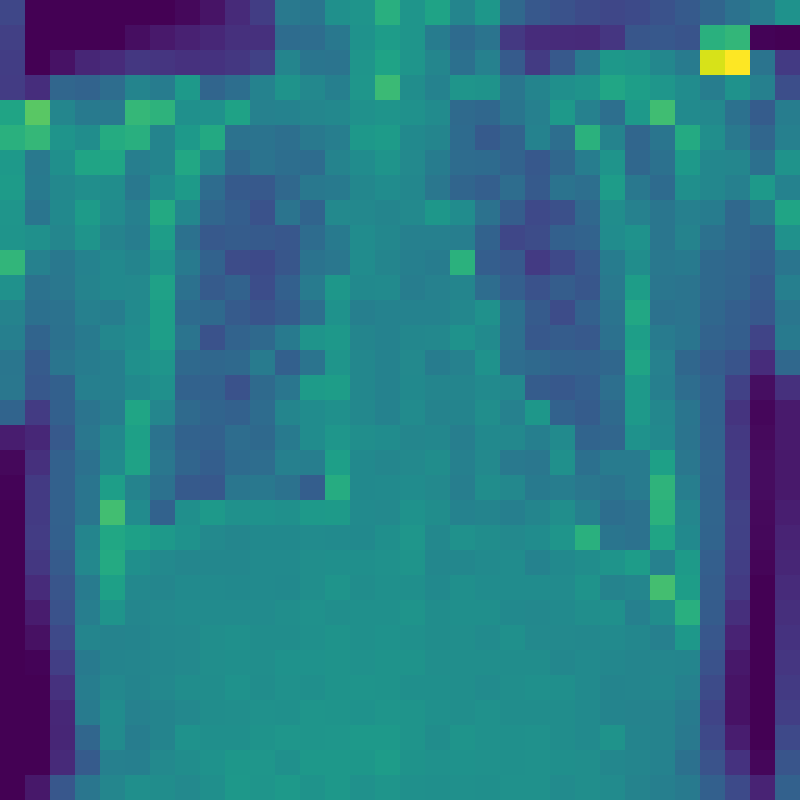}
    \end{minipage}
    \begin{minipage}[t]{0.13\textwidth}
        \centering
        \includegraphics[width=\textwidth]{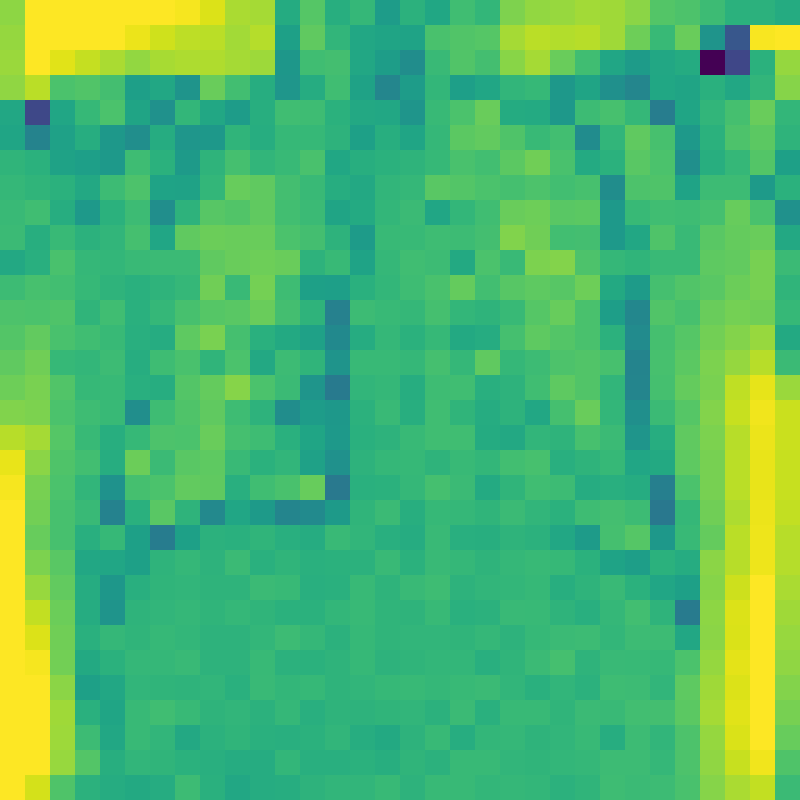}
    \end{minipage}
    \begin{minipage}[t]{0.13\textwidth}
        \centering
        \includegraphics[width=\textwidth]{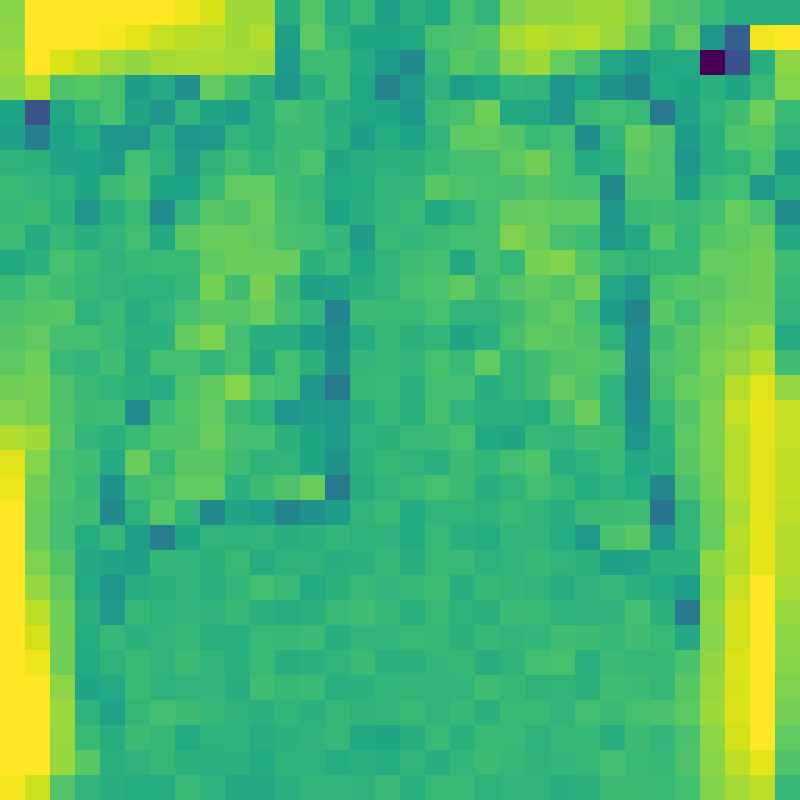}
    \end{minipage}
    \begin{minipage}[t]{0.13\textwidth}
        \centering
        \includegraphics[width=\textwidth]{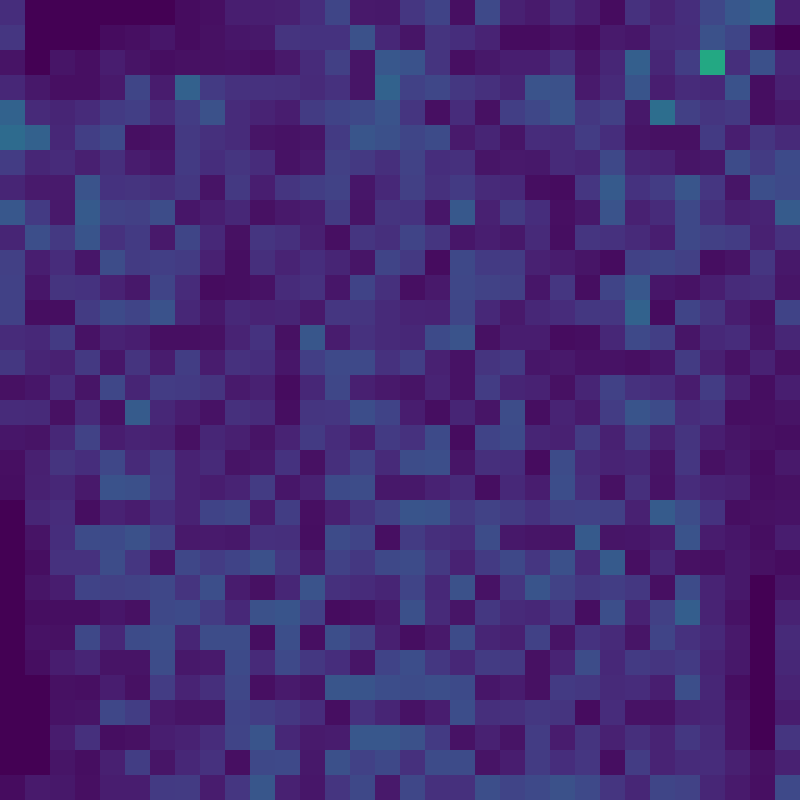}
    \end{minipage}
    \vskip 0.07cm
    \begin{minipage}[t]{0.13\textwidth}
        \centering
        \includegraphics[width=\textwidth]{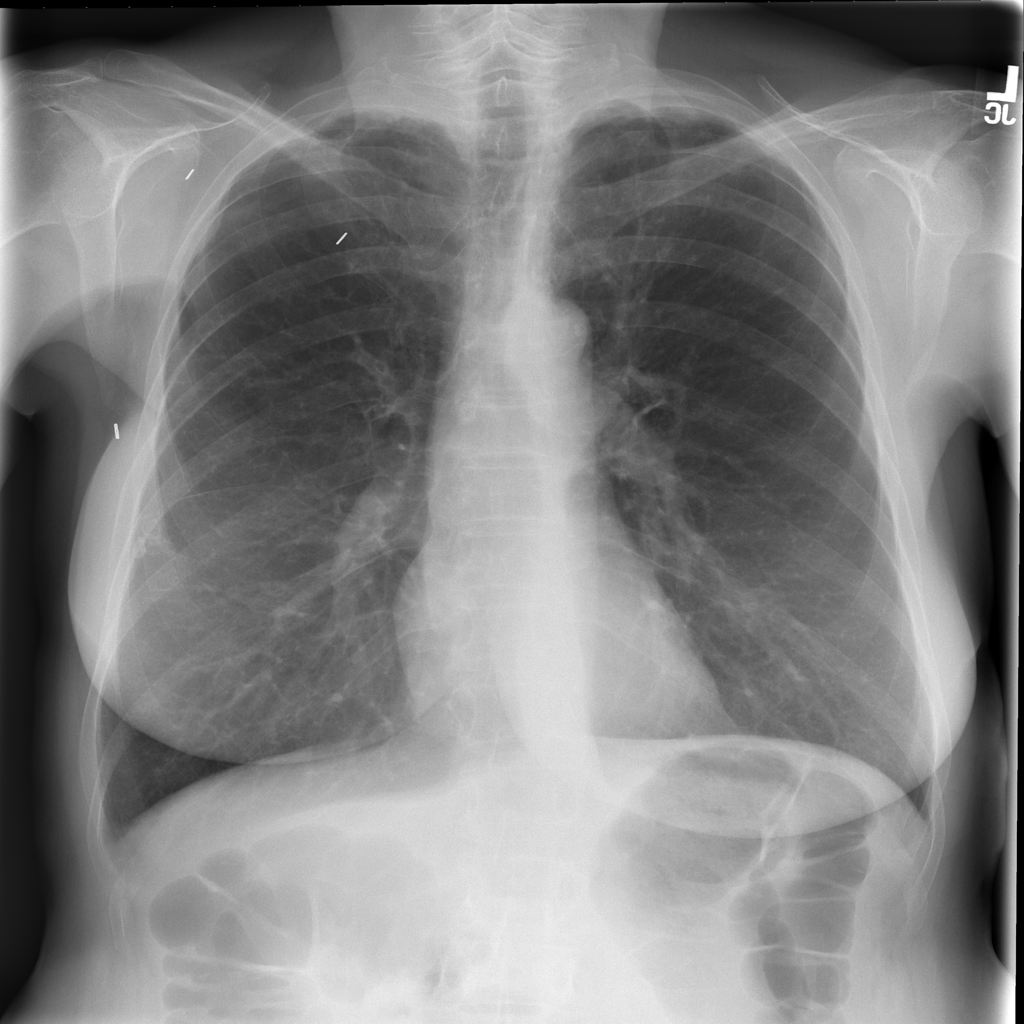}
        \subcaption{}
    \end{minipage}
    \begin{minipage}[t]{0.13\textwidth}
        \centering
        \includegraphics[width=\textwidth]{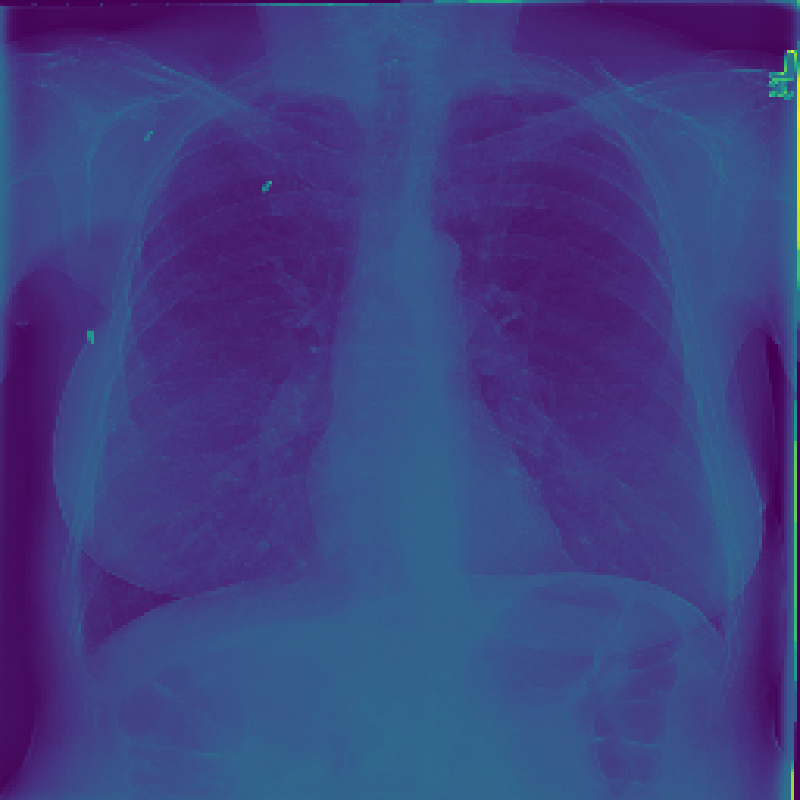}
        \subcaption{}
        \label{d1}
    \end{minipage}
    \begin{minipage}[t]{0.13\textwidth}
        \centering
        \includegraphics[width=\textwidth]{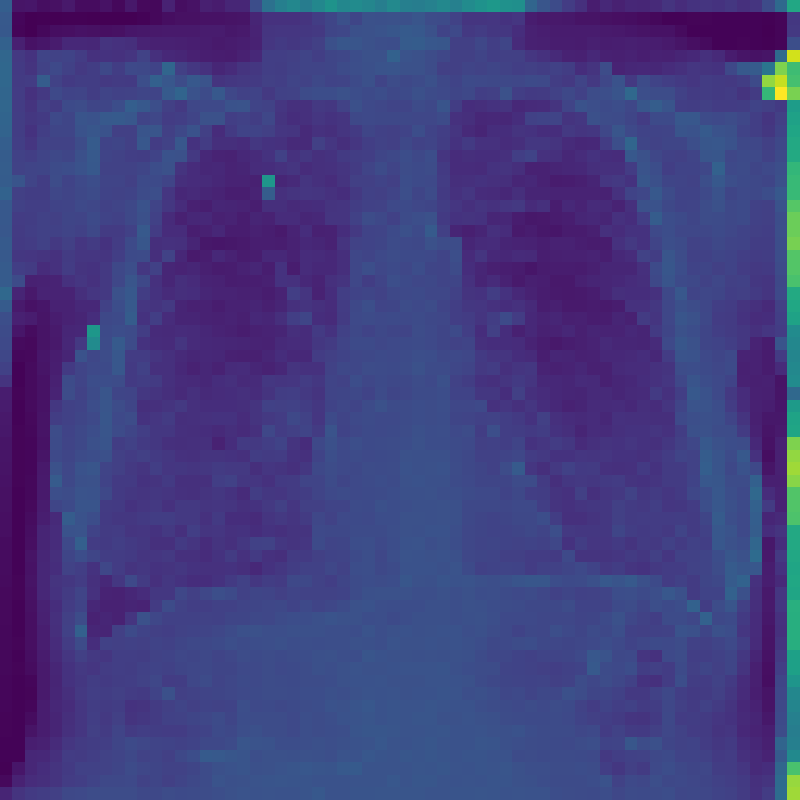}
        \subcaption{}
    \end{minipage}
    \begin{minipage}[t]{0.13\textwidth}
        \centering
        \includegraphics[width=\textwidth]{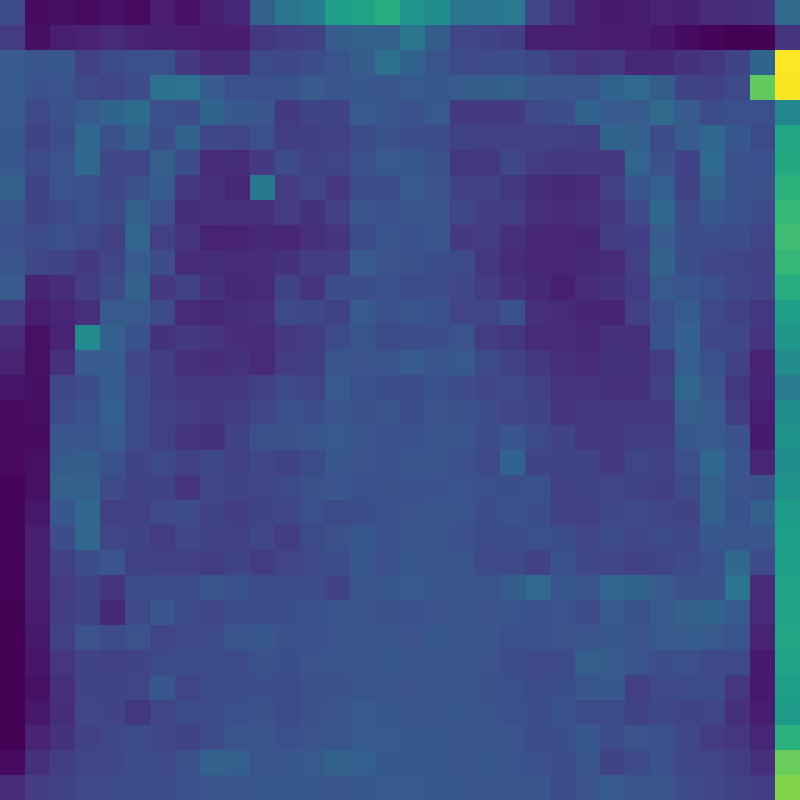}
        \subcaption{}
    \end{minipage}
    \begin{minipage}[t]{0.13\textwidth}
        \centering
        \includegraphics[width=\textwidth]{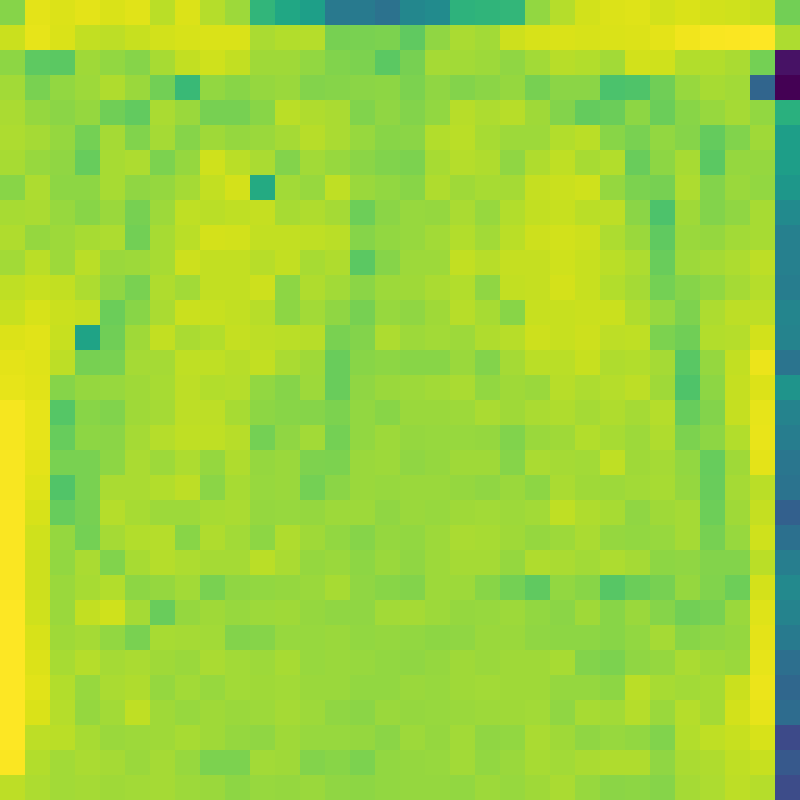}
        \subcaption{}
        \label{d4}
    \end{minipage}
    \begin{minipage}[t]{0.13\textwidth}
        \centering
        \includegraphics[width=\textwidth]{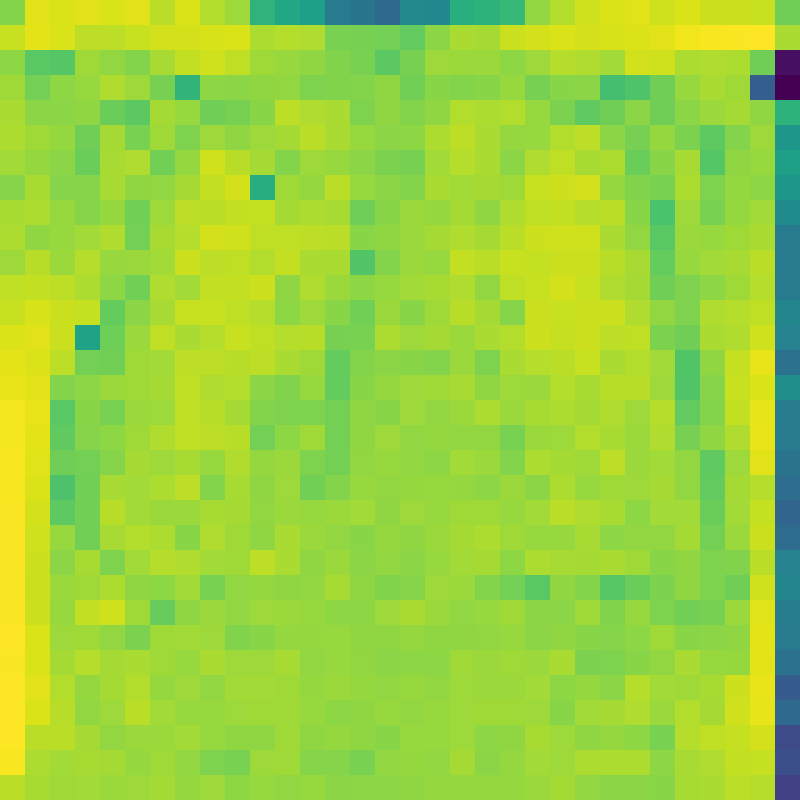}
        \subcaption{}
        \label{qcm}
    \end{minipage}
    \begin{minipage}[t]{0.13\textwidth}
        \centering
        \includegraphics[width=\textwidth]{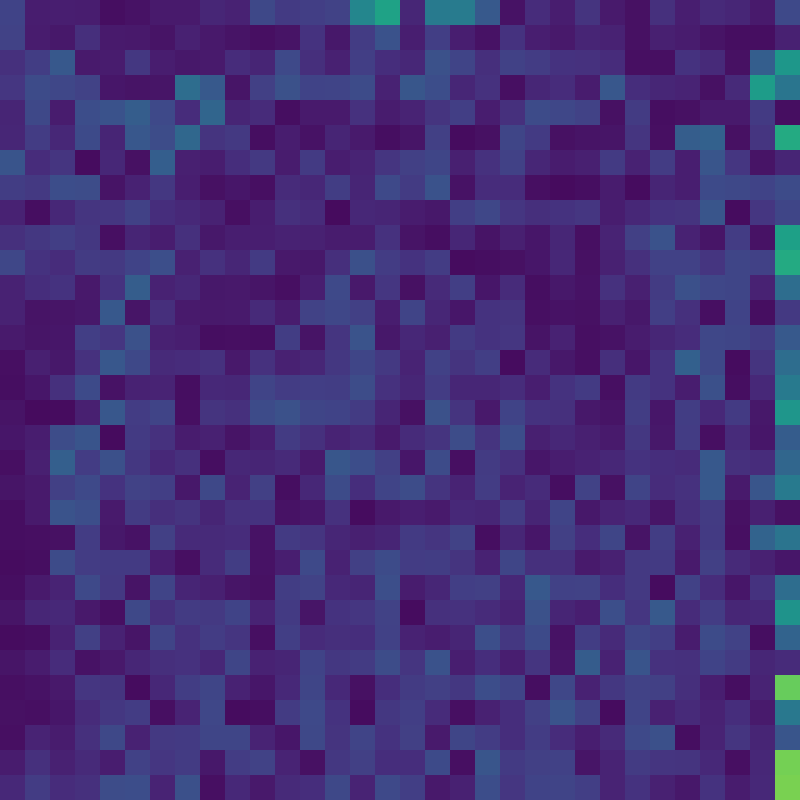}
        \subcaption{}
        \label{res}
    \end{minipage}
    \caption{The feature map visualization of the GMSM and QCM modules. (a) Original image; (b) $f_1$ in Eq. \ref{f}; (c) $f_2$ in Eq. \ref{f}; (d) $f_3$ in Eq. \ref{f}; (e) Output of GMSM $y$; (f) Output of QCM; (g) Residual between $y$ and QCM output.}
    \label{visual}
\end{figure*}
\subsection{Loss Function}
We define the loss function for lossless compression of datasets as follows:
\begin{align}
\mathcal{L}(\theta, G, Q) &= - \mathbb{E}_{x \sim \mathcal{P}_{\text{data}}} \left[ \log_2 \left( p_{\theta} \left( \hat{x} \mid Q(G(x)) \right) \right) \right]  \\
&=-\sum_{k=1}^{N}\sum_{i=1}^{H^{(k)}} \sum_{j=1}^{W^{(k)}} \delta(x_{ij}^{(k)}, 1) \cdot \left(\log_2 p_{\theta} \left( \hat{x}_{ij}^{(k)} \mid Q(G(x^{(k)})) \right) \right)
\end{align}
where $N$ denote the number of images in the dataset, with $W^{(k)}$ and $H^{(k)}$ representing the width and height of the $k$-th image.  The true pixel value at the $i$-th row and $j$-th column of the $k$-th image is one-hot encoded as $x_{ij}^{(k)}$.  $\mathcal{P}_{\text{data}}$ denotes the true data distribution, and $\delta$ is the Kronecker delta function.  The loss function quantifies the storage cost associated with storing the predicted probabilities, and for lossless compression of a single image it's defined as:
\begin{equation}
\mathcal{L}(\theta, G, Q)=-\sum_{i=1}^{H} \sum_{j=1}^{W} \delta(x_{ij}, 1) \cdot \left(\log_2 p_{\theta} \left( \hat{x}_{ij} \mid Q(G(x)) \right) \right)
\end{equation}
\begin{table}[t]
    \centering
    \caption{Comparison of dataset compression performance in BPP. Group 1: Traditional methods, Group 2: Methods based on machine learning.}
    \label{tab1}
    \begin{tabular}{>{\centering\arraybackslash}p{3.0cm}|
                        >{\centering\arraybackslash}p{2.0cm}|
                        >{\centering\arraybackslash}p{2.0cm}|
                        >{\centering\arraybackslash}p{2.0cm}|
                        >{\centering\arraybackslash}p{2.0cm}}
        \hline
        Method & Chest X-ray & CIFAR10 & ImageNet32 & ImageNet64 \\
        \hline
        PNG \cite{b17}       & 3.07  & 5.89  & 6.39  & 5.74  \\
        FLIF \cite{b31}      & 2.87  & 4.19  & 4.52  & 4.54  \\
        JPEG-XL \cite{b32}   & 2.93  & 5.74  & 6.39  & 5.89  \\
        \hline
        L3C \cite{b15}       & 2.89  & -     & 4.76  & 4.42  \\
        IDF \cite{b10}       & 2.78  & 3.34  & 4.18  & 3.90  \\
        Hilloc \cite{b8}     & 3.02  & 4.56  & 4.20  & 3.90  \\
        SHVC \cite{b16}      & 2.83  & 3.16  & 3.98  & 3.68  \\
        iVPF \cite{b12}      & 2.80  & 3.20  & 4.03  & 3.75  \\
        LBB \cite{b9}        & 2.76  & 3.12  & 3.88  & 3.70  \\
        iFlow \cite{b11}     & 2.77  & 3.12  & 3.87  & 3.70  \\
        BCM-Net \cite{b36}   & 2.87  & 3.26  & 4.06  & 4.19  \\
        HMEM \cite{b35}      & 2.79  & 3.19  & 3.89  & 3.80  \\
        ArIB-BPS \cite{b33}  & 2.74  & 3.06  & 3.91  & 3.63  \\
        LVPNet (ours)        &\textbf{2.65$\pm$0.03}  & \textbf{3.02$\pm$0.02}  & \textbf{3.76$\pm$0.04}  & \textbf{3.51$\pm$0.03}  \\
        \hline
    \end{tabular}
\end{table}
\begin{table}[t]
    \centering
    \caption{Comparison of single image compression performance in BPP. $^\Delta$: These values are derived by summing the dataset's compression performance and the initial bit count.}
    \label{tab2}
    \begin{tabular}{>{\centering\arraybackslash}p{3.0cm}|
                        >{\centering\arraybackslash}p{2.0cm}|
                        >{\centering\arraybackslash}p{2.0cm}|
                        >{\centering\arraybackslash}p{2.0cm}|
                        >{\centering\arraybackslash}p{2.0cm}}
        \hline
        Method & Chest X-ray & CIFAR10 & ImageNet32 & ImageNet64 \\
        \hline
        PNG \cite{b17}       & 3.07  & 5.89  & 6.39  & 5.74  \\
        FLIF \cite{b31}      & 2.87  & 4.19  & 4.52  & 4.54  \\
        JPEG-XL \cite{b32}   & 2.93  & 5.74  & 6.39  & 5.89  \\
        \hline
        L3C \cite{b15}       & 2.89  & -     & 4.76  & 4.42  \\
        IDF \cite{b10}       & 2.78  & 3.34  & 4.18  & 3.90  \\
        iVPF \cite{b12}      & 5.83$^\Delta$  & 9.20$^\Delta$  & -  & -  \\
        LBB \cite{b9}        & 37.53$^\Delta$  & 42.98$^\Delta$  & 49.84$^\Delta$  & 41.70$^\Delta$  \\
        iFlow \cite{b11}     & 29.89$^\Delta$  & 37.40$^\Delta$  & 38.27$^\Delta$  & 38.12$^\Delta$  \\
        BCM-Net \cite{b36}   & 3.21  & 4.03  & 4.10  & 4.18  \\
        HMEM \cite{b35}      & 3.15  & 3.97  & 4.04  & 4.13  \\
        ArIB-BPS \cite{b33}  & 2.76  & \textbf{3.07}  & 3.92  & 3.64  \\
        LVPNet (ours)        & \textbf{2.72$\pm$0.02}  & 3.10$\pm$0.02  & \textbf{3.82$\pm$0.03}  & \textbf{3.57$\pm$0.03}  \\
        \hline
    \end{tabular}
\end{table}
\begin{table}[t]
    \centering
    \caption{Comparison of inference times.}
    \label{tab3}
    \begin{tabular}{>{\centering\arraybackslash}p{1.7cm}|
                        >{\centering\arraybackslash}p{2.1cm}|
                        >{\centering\arraybackslash}p{2.0cm}
                        >{\centering\arraybackslash}p{2.0cm}|
                        >{\centering\arraybackslash}p{2.0cm}
                        >{\centering\arraybackslash}p{2.0cm}}
        \hline
         &  &\multicolumn{2}{c|}{Compression Performance(BPP)$\downarrow$}&\multicolumn{2}{c}{Inference Time(ms/sample)$\downarrow$}\\
        \hline
        Dataset & Method & dataset & single & encode & decode\\
        \hline
        \multirow{4}{*}{Chest X-ray}
        &LBB \cite{b9}        & 2.762  & 37.529$^\Delta$  & 643.95  & 634.19  \\
        &iFlow \cite{b11}     & 2.768  & 29.894$^\Delta$  & 451.87  & 512.48  \\
        &ArIB-BPS \cite{b33}  & 2.741  & 2.763  & 197.24  & 203.50  \\
        &LVPNet (ours)        & \textbf{2.650}  & \textbf{2.719}  & \textbf{143.63}  & \textbf{184.06}  \\
        \hline
        \multirow{4}{*}{CIFAR10}
        &LBB \cite{b9}        & 3.118  & 49.835  & 64.94  & 64.94  \\
        &iFlow \cite{b11}     & 3.118  & 37.398  & 19.38  & 47.56  \\
        &ArIB-BPS \cite{b33}  & 3.057  & \textbf{3.070}  & \textbf{14.93}  & \textbf{14.93}  \\
        &LVPNet (ours)        & \textbf{3.024}  & 3.099  & 16.44  & 20.28  \\
        \hline
        \multirow{4}{*}{ImageNet32}
        &LBB \cite{b9}        & 3.875  & 49.835  & 194.14  & 194.14  \\
        &iFlow \cite{b11}     & 3.873  & 38.273  & 74.84  & 119.30  \\
        &ArIB-BPS \cite{b33}  & 3.911  & 3.918  & 52.96  & 52.96  \\
        &LVPNet (ours)        & \textbf{3.758}  & \textbf{3.822}  & \textbf{39.73}  & \textbf{48.17}  \\
        \hline
    \end{tabular}
\end{table}
\section{Experiments}
\subsection{Training Details}
We evaluate the performance of our method for both dataset-level and single-image lossless compression on Chest X-ray\cite{b30}, CIFAR10, ImageNet32, and ImageNet64. Of these, Chest X-ray is a medical image dataset, while the others are natural image datasets. We use bits per pixel (BPP) as the evaluation metric, where lower values indicate better performance. 

We use the Adam optimizer \cite{b29} with an initial learning rate of 0.0001 and step decay for scheduling. The GMSM sampling rate $r$ is set to 0.15, and the quantization step size $Q_{step}$ is 0.01. For dataset compression, the QCM consists of 24 residual blocks, while for single image compression, it contains only 3 residual blocks. All experiments are conducted on an NVIDIA GeForce RTX 3060.
\subsection{Visualization of Feature Maps}
In this section, we visualize the feature maps of the GMSM and QCM modules, as shown in Figure \ref{visual}. From Figures \ref{d1} to \ref{d4}, we observe that, during GMSM's downsampling process, the earlier layers effectively capture the skeletal and organ contours in the X-ray image, while the later layers extract more abstract features. This observation suggests that GMSM progressively extracts key information in a coarse-to-fine manner. As shown in Figure \ref{qcm}, the output from the QCM module closely resembles the feature map of $y$ in Figure \ref{d4}. Additionally, the residual map in Figure \ref{res} appears generally darker, indicating the QCM module's effectiveness in compensating for quantization loss.
\subsection{Compression Performance}
We compare our method with traditional approaches and state-of-the-art deep learning methods. We present the detailed results of the experiment in Table \ref{tab1}. Our method achieves superior compression performance, particularly in compressing medical image datasets, while also maintaining competitive performance on natural image datasets. ArIB-BPS shows limited compression capability due to issues like posterior collapse and inefficient use of latent variables, highlighting GMSM's effectiveness in leveraging global latent variables for prediction.

To further validate the effectiveness of our method, we evaluate its performance on the lossless compression of individual images, with results presented in Table \ref{tab2}. The experimental findings indicate that our method achieves high compression efficiency in single-image compression tasks. Compared to state-of-the-art lossless compression methods, it consistently outperforms existing approaches across multiple datasets. These results further demonstrate the applicability and robustness of our method, proving its suitability for both large-scale dataset compression and single-image lossless compression.

In addition, we perform a comprehensive comparison between our method and state-of-the-art approaches, which demonstrate strong performance in both large-scale image datasets and single-image compression tasks. As shown in Table \ref{tab3}, LVPNet achieves outstanding performance across various datasets while achieving an inference speed on par with the fastest existing approach, indicating a better balance between compression efficiency and inference speed.
\subsection{Ablation Studies}
\begin{table}[t]
    \centering
    \caption{Ablation study of the effectiveness of GMSM and QCM.}
    \label{tab4}
    \begin{tabular}{>{\centering\arraybackslash}p{1.7cm}
                        >{\centering\arraybackslash}p{1.4cm}|
                        >{\centering\arraybackslash}p{2.0cm}
                        >{\centering\arraybackslash}p{2.0cm}|
                        >{\centering\arraybackslash}p{2.0cm}
                        >{\centering\arraybackslash}p{2.0cm}}
        \hline
          &  &\multicolumn{2}{c|}{Compression Performance(BPP)$\downarrow$}&\multicolumn{2}{c}{Inference Time(ms/sample)$\downarrow$}\\
        \hline
        Sampling & QCM & dataset & single & encode & decode\\
        \hline
        CNN &  & 2.833 & 2.906 & 133.11 & \textbf{166.89}\\
        CNN & \checkmark & 2.757 & 2.859 & \textbf{132.15} & 184.12\\
        GMSM &  & 2.732 & 2.841 & 143.52 & 167.03\\
        GMSM & \checkmark & \textbf{2.650} & \textbf{2.719} & 143.63 & 184.06\\
        \hline
    \end{tabular}
\end{table}
\begin{figure}[!t]
    \centering
    \begin{subfigure}[b]{0.49\textwidth}
        \centering
        \includegraphics[width=\textwidth]{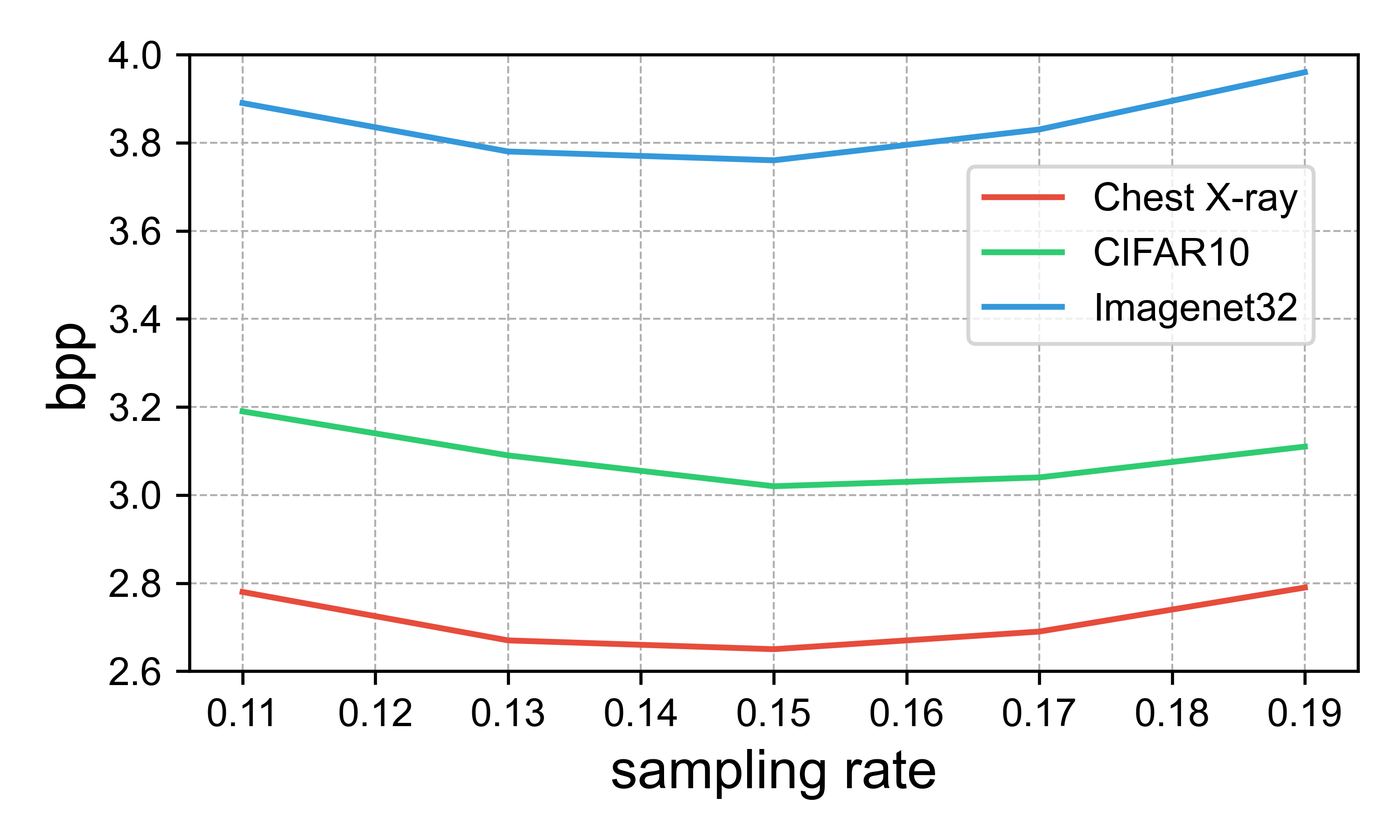}
        \caption{Dataset compression}
        \label{fig:subfig1}
    \end{subfigure}
    \hfill
    \begin{subfigure}[b]{0.49\textwidth}
        \centering
        \includegraphics[width=\textwidth]{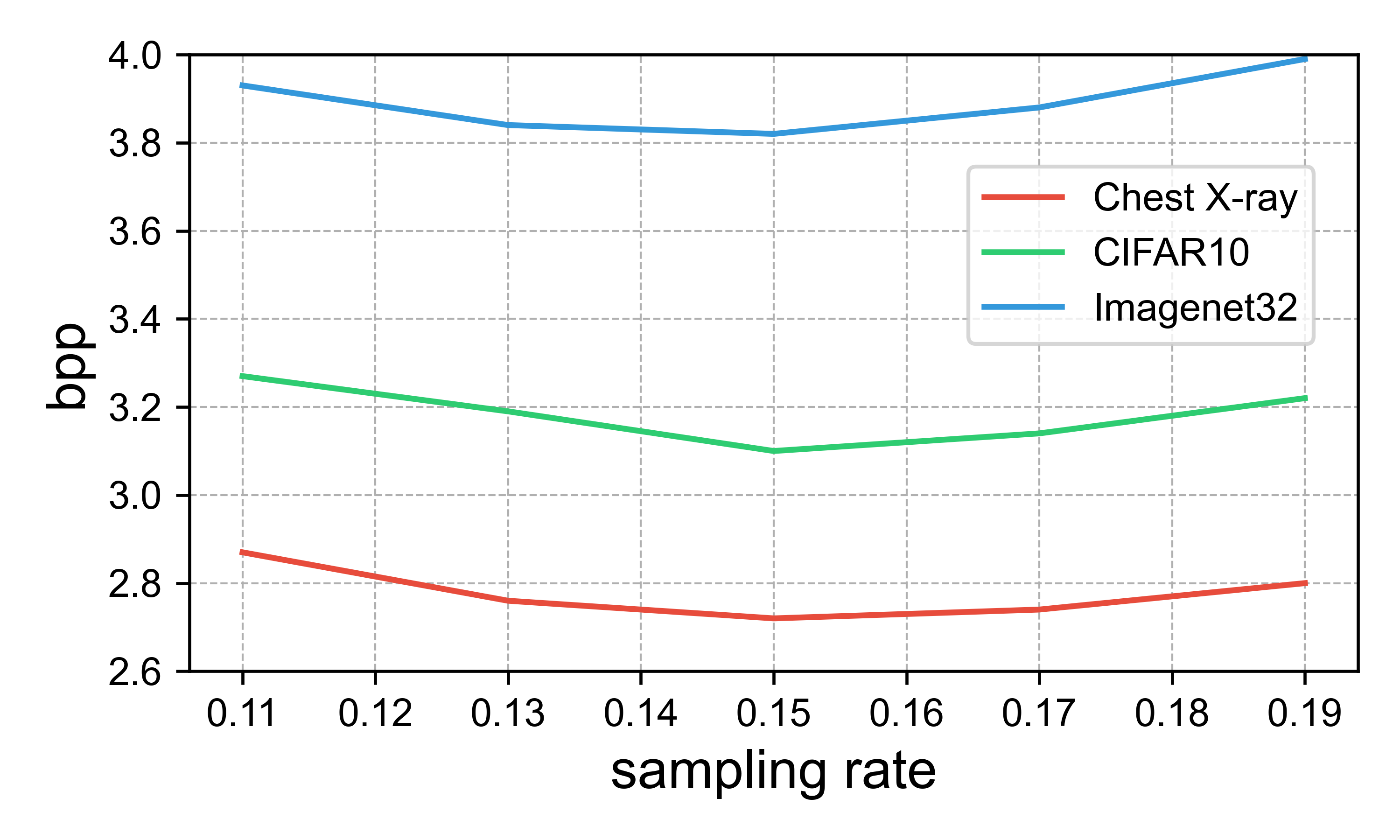}
        \caption{Single image compression}
        \label{fig:subfig2}
    \end{subfigure}
    \caption{Ablation study for different sampling rates $r$.}
    \label{sample}
\end{figure}
We perform three ablation experiments on the Chest X-ray dataset to demonstrate the efficacy of our latent-variable-based, prediction-driven model for medical image lossless compression and the effectiveness of discrete sampling. We assess the impact of autoregression across different dimensions by training and evaluating three additional models: one in which the global multi-level perception module is replaced by a Convolutional Neural Network (CNN), one in which the quantization compensation network is removed, and one where both modifications are applied.
 As shown in the experimental results in Table \ref{tab4}, our model achieves superior compression performance while maintaining comparable inference time. Additionally, we evaluate the effectiveness of our customized sampling rate strategy. As shown in Figure \ref{sample}, this sampling rate provides better compression performance, striking a favorable balance between latent variable storage overhead and compression prediction accuracy.
\section{Conclusion}
We propose a novel method for lossless medical image compression, LVPNet. The proposed GMSM module allows us to make full use of latent variables, effectively modeling the spatial dependencies within the latent space. In addition, we introduce a quantization compensation module, which corrects quantized features by modeling the distribution of quantization errors, thereby mitigating the effects of quantization loss. Experimental results demonstrate that, compared to the current state-of-the-art methods in lossless image compression, our method achieves superior compression performance with comparable inference time. 


\begin{thebibliography}{8}
\bibitem{b38}
Li, Junjian, et al. "CA 2 CL: Cluster-Aware Adversarial Contrastive Learning for Pathological Image Analysis." IEEE Journal of Biomedical and Health Informatics (2025).

\bibitem{b37}
Xu, Jiawei, et al. "Semantic-Orthogonal Multi-modal Attention Network for RGB-D Salient Object Detection." The Visual Computer (2025): 1-13.

\bibitem{b39}
Li, Junjian, et al. "DARC: Deep adaptive regularized clustering for histopathological image classification." Medical image analysis 80 (2022): 102521.

\bibitem{b50}
Qiu, Kunpeng, et al. "Noise-Consistent Siamese-Diffusion for Medical Image Synthesis and Segmentation." Proceedings of the Computer Vision and Pattern Recognition Conference. 2025.

\bibitem{b51}
Qiu, Kunpeng, Zhiying Zhou, and Yongxin Guo. "Learn From Zoom: Decoupled Supervised Contrastive Learning For WCE Image Classification." ICASSP 2024-2024 IEEE International Conference on Acoustics, Speech and Signal Processing (ICASSP). IEEE, 2024.

\bibitem{b1}
Shannon, Claude Elwood. "A mathematical theory of communication." The Bell system technical journal 27.3 (1948): 379-423.

\bibitem{b17}
Boutell, Thomas. Png (portable network graphics) specification version 1.0. No. rfc2083. 1997.

\bibitem{b18}
Ziv, Jacob, and Abraham Lempel. "A universal algorithm for sequential data compression." IEEE Transactions on information theory 23.3 (1977): 337-343.

\bibitem{b19}
Huffman, David A. "A method for the construction of minimum-redundancy codes." Proceedings of the IRE 40.9 (1952): 1098-1101.

\bibitem{b20}
"JPEG 2000 standard ISO/IEC 15444 ITU-T Recommendation T.800", Joint Photographic Experts Group (JPEG), 2004, [online] Available: http://www.jpeg.org/jpeg.

\bibitem{b2}
Bond-Taylor, Sam, et al. "Deep generative modelling: A comparative review of vaes, gans, normalizing flows, energy-based and autoregressive models." IEEE transactions on pattern analysis and machine intelligence 44.11 (2021): 7327-7347.

\bibitem{b40}
Yu, Xinlei, et al. "ICH-PRNet: a cross-modal intracerebral haemorrhage prognostic prediction method using joint-attention interaction mechanism." Neural Networks 184 (2025): 107096.

\bibitem{b41}
Luo, Haozhe, et al. "On the Interplay of Human-AI Alignment, Fairness, and Performance Trade-offs in Medical Imaging." arXiv preprint arXiv:2505.10231 (2025).

\bibitem{b3}
Salimans, Tim, et al. "Pixelcnn++: Improving the pixelcnn with discretized logistic mixture likelihood and other modifications." arXiv preprint arXiv:1701.05517 (2017).

\bibitem{b4}
Van Den Oord, Aäron, Nal Kalchbrenner, and Koray Kavukcuoglu. "Pixel recurrent neural networks." International conference on machine learning. PMLR, 2016.

\bibitem{b33}
Zhang, Zhe, et al. "Learned Lossless Image Compression based on Bit Plane Slicing." CVPR. 2024.

\bibitem{b5}
Kingma, Friso, Pieter Abbeel, and Jonathan Ho. "Bit-swap: Recursive bits-back coding for lossless compression with hierarchical latent variables." International Conference on Machine Learning. PMLR, 2019.

\bibitem{b6}
Mentzer, Fabian, et al. "Practical full resolution learned lossless image compression." CVPR. 2019.

\bibitem{b7}
Townsend, James, Tom Bird, and David Barber. "Practical lossless compression with latent variables using bits back coding." arXiv preprint arXiv:1901.04866 (2019).

\bibitem{b8}
Townsend, James, et al. "Hilloc: Lossless image compression with hierarchical latent variable models." arXiv preprint arXiv:1912.09953 (2019).

\bibitem{b34}
Bai, Yuanchao, et al. "Deep lossy plus residual coding for lossless and near-lossless image compression." TPAMI 46.5 (2024): 3577-3594.

\bibitem{b35}
Fu, Chuan, Bo Du, and Liangpei Zhang. "Hybrid-context-based multi-prior entropy modeling for learned lossless image compression." Pattern Recognition 155 (2024): 110632.

\bibitem{b9}
Ho, Jonathan, Evan Lohn, and Pieter Abbeel. "Compression with flows via local bits-back coding." Advances in Neural Information Processing Systems 32 (2019).

\bibitem{b10}
Hoogeboom, Emiel, et al. "Integer discrete flows and lossless compression." Advances in Neural Information Processing Systems 32 (2019).

\bibitem{b11}
Zhang, Shifeng, et al. "iflow: Numerically invertible flows for efficient lossless compression via a uniform coder." Advances in Neural Information Processing Systems 34 (2021): 5822-5833.

\bibitem{b12}
Zhang, Shifeng, et al. "ivpf: Numerical invertible volume preserving flow for efficient lossless compression." CVPR. 2021.

\bibitem{b21}
Ulacha, Grzegorz, and Mirosław Łazoryszczak. "Lossless Image Compression Using Context-Dependent Linear Prediction Based on Mean Absolute Error Minimization." Entropy 26.12 (2024): 1115.

\bibitem{b22}
Rhee, Hochang, et al. "Lossless image compression by joint prediction of pixel and context using duplex neural networks." IEEE Access 9 (2021): 86632-86645.

\bibitem{b23}
Rahman, Md Atiqur, and Mohamed Hamada. "A prediction-based lossless image compression procedure using dimension reduction and Huffman coding." Multimedia Tools and Applications 82.3 (2023): 4081-4105.

\bibitem{b36}
Liu, Xiangrui, et al. "Bilateral context modeling for residual coding in lossless 3D medical image compression." TIP (2024).

\bibitem{b13}
Hoogeboom, Emiel, et al. "Autoregressive diffusion models." arXiv preprint arXiv:2110.02037 (2021).

\bibitem{b14}
Kingma, Diederik, et al. "Variational diffusion models." Advances in neural information processing systems 34 (2021): 21696-21707.

\bibitem{b15}
Mentzer, Fabian, et al. "Practical full resolution learned lossless image compression." CVPR. 2019.

\bibitem{b16}
Ryder, Tom, et al. "Split hierarchical variational compression." CVPR. 2022.

\bibitem{b44}
Yang, Yanwu, et al. "Creg-kd: Model refinement via confidence regularized knowledge distillation for brain imaging." Medical Image Analysis 89 (2023): 102916.

\bibitem{b25}
Lucas, James et al. “Don't Blame the ELBO! A Linear VAE Perspective on Posterior Collapse.” ArXiv abs/1911.02469 (2019): n. pag.

\bibitem{b42}
Luo, Haozhe, Yu Changdong, and Raghavendra Selvan. "Hybrid ladder transformers with efficient parallel-cross attention for medical image segmentation." International conference on medical imaging with deep learning. PMLR, 2022.

\bibitem{b43}
Yang, Yanwu, et al. "Advancing Brain Imaging Analysis Step-by-Step via Progressive Self-paced Learning." International Conference on Medical Image Computing and Computer-Assisted Intervention. Cham: Springer Nature Switzerland, 2024.

\bibitem{b26}
He, Kaiming, et al. "Deep residual learning for image recognition." CVPR. 2016.

\bibitem{b27}
Huffman, David A. "A method for the construction of minimum-redundancy codes." Proceedings of the IRE 40.9 (1952): 1098-1101.

\bibitem{b28}
B. P. Tunstall, “Synthesis of noiseless compression codes,” in PhD
 diss., Georgia Institute of Technology, 1967.

\bibitem{b31}
Sneyers, Jon, and Pieter Wuille. "FLIF: Free lossless image format based on MANIAC compression." ICIP. IEEE, 2016.

\bibitem{b32}
Alakuijala, Jyrki, et al. "JPEG XL next-generation image compression architecture and coding tools." Applications of digital image processing XLII. Vol. 11137. SPIE, 2019.

\bibitem{b30}
Wang, Xiaosong, et al. "Chestx-ray8: Hospital-scale chest x-ray database and benchmarks on weakly-supervised classification and localization of common thorax diseases." CVPR. 2017.

\bibitem{b29}
Kingma, Diederik P. and Jimmy Ba. “Adam: A Method for Stochastic Optimization.” CoRR abs/1412.6980 (2014): n. pag.

\end{thebibliography}
\end{document}